\documentclass[]{interact}

\usepackage{epstopdf}
\usepackage[caption=false]{subfig}

\usepackage[numbers,sort&compress]{natbib}
\bibpunct[, ]{[}{]}{,}{n}{,}{,}
\renewcommand\bibfont{\fontsize{10}{12}\selectfont}

\theoremstyle{plain}

\theoremstyle{definition}

\theoremstyle{remark}

\begin{document}


\title{Machine Learning Methods for \\ Spectral Efficiency Prediction  in Massive MIMO Systems}

\author{
\name{Evgeny~Bobrov\textsuperscript{a,c,*}\thanks{CONTACT Evgeny~Bobrov. Email: eugenbobrov@ya.ru}, Sergey~Troshin\textsuperscript{b,*}, Nadezhda~Chirkova\textsuperscript{b},  Ekaterina~Lobacheva\textsuperscript{b}, Sviatoslav~Panchenko\textsuperscript{c, e}, Dmitry~Vetrov\textsuperscript{b, d}, Dmitry~Kropotov\textsuperscript{a, b}}
\affil{\textsuperscript{a}Lomonosov MSU, Russia; \textsuperscript{b}HSE University, Russia, \textsuperscript{c}MRC, Huawei Technologies, Russia, \textsuperscript{d}AIRI, Russia, \textsuperscript{e}MIPT, Russia, \textsuperscript{*}Equal contribution}}

\maketitle

\begin{abstract}
Channel decoding, channel detection, channel assessment, and resource management for wireless multiple-input multiple-output (MIMO) systems are all examples of problems where machine learning (ML) can be successfully applied. In this paper, we study several ML approaches to solve the problem of estimating the spectral efficiency (SE) value for a certain precoding scheme, preferably in the shortest possible time. The best results in terms of mean average percentage error (MAPE) are obtained with gradient boosting over sorted features, while linear models demonstrate worse prediction quality. Neural networks perform similarly to gradient boosting, but they are more resource- and time-consuming because of hyperparameter tuning and frequent retraining.  We investigate the practical applicability of the proposed algorithms in a wide range of scenarios generated by the Quadriga simulator. In almost all scenarios, the MAPE achieved using gradient boosting and neural networks is less than 10\%.
\end{abstract}

\begin{keywords}
MIMO, Precoding, Machine Learning,  Gradient Boosting, Neural Networks.
\end{keywords}

\section{Introduction}

Wireless multiple-input multiple-output (MIMO) technology is the subject of extensive theoretical and practical research for next-generation cellular systems \cite{wannstrom2013lte, huh2012achieving}, which consider multiuser MIMO as one of the core technologies \cite{huh2011network}. A considerable research effort has been dedicated to performance evaluation of MIMO systems in realistic cellular environments \cite{farajidana20093gpp}. One of the most important features in mobile systems is to provide much higher rate data services to a large number of users without corresponding increase in transmitter power and bandwidth \cite{wang2013spectral}. The efficiency of communication systems is traditionally measured in terms of spectral efficiency (SE), which is directly related to the channel capacity in bit/s. This metric indicates how efficiently a limited frequency spectrum is utilized \cite{heliot2012energy}.  

Evaluation of the channel capacity for the MIMO system in terms of SE has attracted considerable research interest in the past decades \cite{foschini1998limits, telatar1999capacity}. The cell averaged SE is an important parameter for evaluating performance of cellular systems, and it is often obtained by using sophisticated system-level simulations. For traditional cellular systems, the cell-wide SE was studied \cite{alouini1999area}. The problem of user scheduling in 5G systems requires fast SE computations \cite{chataut2019channel, liu2016efficient}, which can be enabled using machine learning (ML) methods \cite{ullah2020machine}.

Advanced ML techniques are capable of providing simple and efficient solutions, given that complicated phases of design and training are completed. ML has recently been applied to power-control problems in wireless communications. In \cite{sun2017learning} a deep neural network for optimizing the averaging SE is constructed for a system serving a few dozens of users. The neural network structure can be reused \cite{zappone2018model} for solving an energy-efficiency problem. In \cite{van2019sum} the joint data and pilot non-convex power optimization for maximum SE in multi-cell Massive MIMO systems is studied. In \cite{sanguinetti2018deep} deep learning (DL) is used to solve max-min and max-prod power allocation (PA) problems in downlink of Massive MIMO. In \cite{zhao2020power, d2019uplink} the PA problem in cell-free Massive MIMO is solved using DL, which is closely related to the SE evaluation.

Recently the application of neural networks to massive MIMO got significant attention in the literature. Neural networks can be applied to channel decoding, detection, channel estimation, and resource management \cite{xia20}. A supervised neural-network-based approach for precoding (predicting precoding matrix $W$ using the dataset of (channel matrix $H$, precoding matrix $W$) pairs) for the multiple-input and single-output (MISO) systems is proposed. Also, a mixed problem statement is considered where precoding is computed via a conventional iterative scheme, while the neural network predicts optimal per-user power allocation, which is utilized in the iterative scheme. In \cite{huang20} an unsupervised neural-network-based approach for precoding in a MIMO system is proposed, where a neural network predicts the precoding based on a channel matrix, and spectral efficiency is used as the loss function directly, so there is no need to provide precoding as targets during training. There are several attempts \cite{guo2020regression, ullah2020machine, rozenblit2018machine} to predict signal to interference-and-noise-ratio (SINR), which is closely related to predicting SE, but these studies are very limited and take into account only the power distributions of users, but not a channel, precoding, detection matrices.

Although the aforementioned studies deal with problem statements that are somewhat comparable to the problem we solve, there is still a lot of potential for research. Firstly, these works mostly do not describe data generation, so it is unclear whether their results are applicable in practice. In contrast, we study the applicability of ML models in a wider range of Quadriga scenarios. Secondly, the previous research only describes a basic approach for solving the problem and does not study the influence of different neural network architectures, i. e. transformers, and does not consider other efficient machine learning algorithms, e. g. gradient boosting or linear models, that may contribute a lot to the quality of the solution. We include the mentioned algorithms and architectures in our experiments. Finally, all of the works above do not focus on SE prediction, while we consider it as one of our goals. To our knowledge, the prediction of SE using ML for a detailed MIMO model has not been previously studied.

The remainder of the paper is organized as follows.  In Section~\ref{sec:MIMO Background} we describe a Massive MIMO system model, particular precoding methods, quality measures, types of detection matrices, and power constraints. In Section~\ref{sec:SE Problem} we consider the problem of spectral efficiency estimation using machine learning methods. In Section~\ref{sec:se_approach} we describe channel dataset, algorithm features, and standard machine learning approaches including linear models, neural networks, and gradient boosting algorithms. Section~\ref{sec:experiments} contains numerical results and Section~\ref{sec:conclusion} concludes the paper. Appendix Section~\ref{sec:appendix} contains results for a transformer-based method.

\section{MIMO System Background}\label{sec:MIMO Background}

\begin{figure}
  \centering
  \includegraphics[scale=0.8]{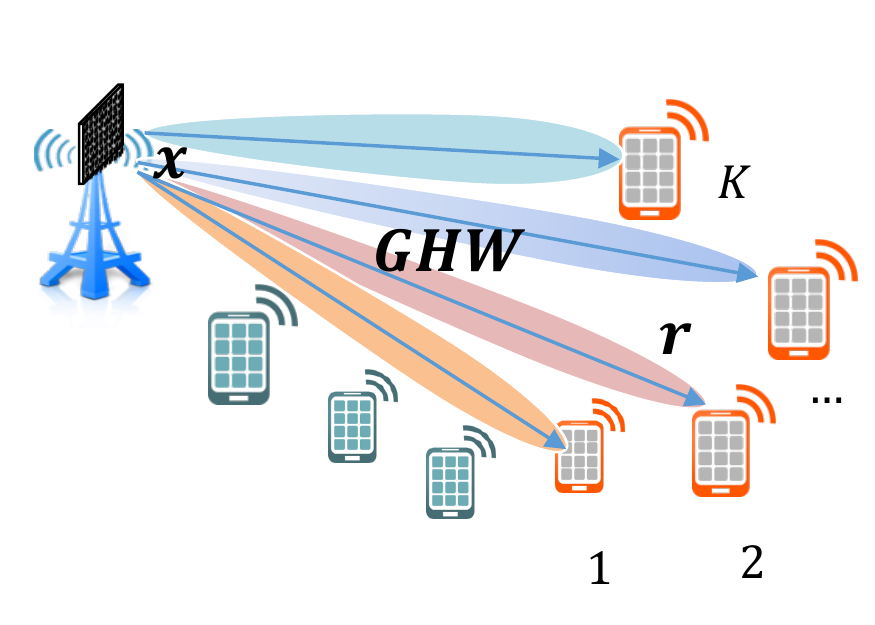}
  \includegraphics[width=\linewidth]{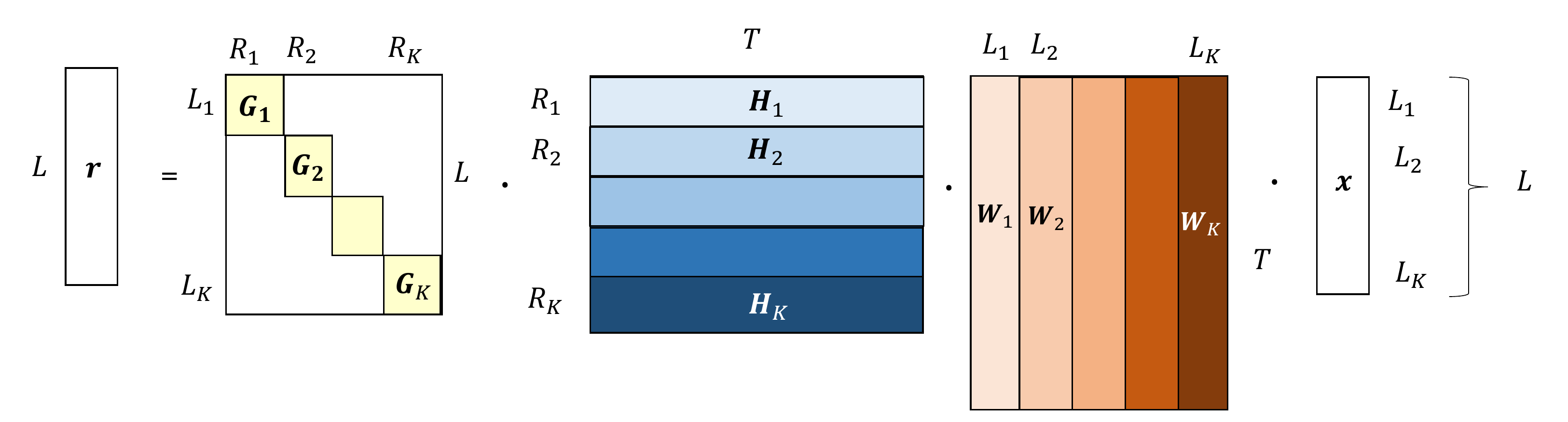}
  \caption{System model. Multi-User Precoding allows to transmit different information to different users simultaneously. The problem is to predict spectral efficiency function for a given precoding matrix $W$.}
  \label{fig:system_model}
\end{figure}

We consider a precoding problem in multi-user massive MIMO communication in 5G cellular networks. In such system, a base station has multiple transmitting antennas that emit signals to several users simultaneously. Each user also has multiple receiving antennas. The base station measures the quality of channel between each transmitter and receiver. This is known as channel state information. The precoding problem is to find an appropriate weighting (phase and gain) of the signal transmission in such a way that the spectral efficiency is maximum at the receiver. We consider the following downlink multi-user linear channel:
\begin{equation}\label{eq:system_model}
    r_k = G_k ( H_k W x + n_k), \quad  k = 1 \dots K.
\end{equation}
In this model, we have $K$ users and we would like to transmit $L_k$ symbols to $k$-th user. Hence in total we would like to transmit a vector $x\in\mathbb{C}^L$, where $L=L_1+\dots+L_K$. First, we multiply the vector being transmitted by a precoding matrix $W\in\mathbb{C}^{T{\times}L}$, where $T$ is the total number of transmitting antennas on the base station. Then we transmit the precoded signal to all users. Suppose that $k$-th user has $R_k$ receiving antennas and $H_k\in\mathbb{C}^{R_k{\times}T}$ is a channel between $k$-th user antennas and antennas on the base station. Then $k$-th user receives $H_kWx + n_k$, where $n_k$ is Gaussian noise. Finally, $k$-th user applies a detection for transmitted symbols by multiplying the received vector by a detection matrix $G_k\in\mathbb{C}^{L_k{\times}R_k}$. The vector of detected symbols for the $k$-th users is denoted by $r_k\in\mathbb{C}^{L_k}$. The whole process of symbol transmission is presented in Fig.~\ref{fig:system_model} \cite{Conjugate}. Usually, in downlink the numbers of symbols being transmitted, user antennas, and base station antennas are related as $L_k \leqslant R_k \leqslant T$.


\subsection{Precoding Methods}

We denote a concatenation of user channel matrices as $H=[H_1,\dots,H_K]\in\mathbb{C}^{R{\times}T}$. Then we make a singular-value decomposition of each matrix as $H_k=U_k^\mathrm{H}S_kV_k$, where $U_k\in\mathbb{C}^{R_k{\times}R_k}$ is a unitary matrix, $S_k\in\mathbb{R}^{R_k{\times}R_k}$ is a diagonal matrix and $V_k \in\mathbb{C}^{R_k{\times}T}$ is a semi-unitary matrix. In such way we obtain the decomposition: $H=[U_1^\mathrm{H}S_1V_1,\dots,U_K^\mathrm{H}S_KV_K]\in\mathbb{C}^{R{\times}T}$. For each $V_k$ we denote a sub-matrix $\widetilde{V}_k \in \mathbb C ^ {L_k{\times}T}$ with rows corresponding to the $L_k$ largest singular values from $S_k$. We denote a concatenation of all $\widetilde{V}_k$ as $\widetilde{V}=[\widetilde{V}_1,\dots,\widetilde{V}_K]\in\mathbb{C}^{L{\times}T}$ .

There are some well-known heuristic precoding algorithms \cite{MRT, RZF, ZF, RZF2, RZF19}:
\begin{align}
    &W_{\mathrm{MRT}} = \mu \widetilde{V}^\mathrm{H}P \in \mathbb C^{T \times L} \mbox{ -- Maximum Ratio Transmission (MRT)}, \\
    &W_{\mathrm{ZF}} = \mu \widetilde{V}^\mathrm{H}(\widetilde{V}\widetilde{V}^\mathrm{H})^{-1}P \in \mathbb C ^ {T \times L} \mbox{ -- Zero-Forcing (ZF)}.
\end{align}
The diagonal matrix $P\in\mathbb{C}^{L{\times}L}$ is a column-wise power normalization of precoding matrix, $\mu $ is a scalar power normalization constant for meeting per-antenna power constraints~\eqref{eq:power_constraints}. Normalization by constant $\mu$ is done as the last step, after normalization by $P$. The value $\sigma^2$ is a variance of Gaussian noise $n_k$ during transmission \eqref{eq:system_model}. We also consider LBFGS precoding optimization scheme \cite{bobrov2021study}.  



\subsection{Quality Functions}

The base station chooses an optimal precoding matrix $W$ based on measured channel $H$, which maximizes the so-called \textit{Spectral Efficiency} (\textit{SE}) \cite{SE, Gotoh01, Benson06}. Function of SE is closely related to \textit{Signal-to-Interference-and-Noise Ratio} and is expressed as :
\begin{equation}
    \mathrm{SE}(W, H, G, \sigma^2) =  \frac{1}{K} \sum_{k=1}^K L_k \log_2 (1 + \mathrm{SINR}_k^{eff}(W, H_k, G_k, \sigma^2)).
    \label{eq:spectral_efficiency}
\end{equation}
The set of symbol indexes targeted to $k$-th user is denoted as $\mathcal{L}_k$:
\begin{align}
    &\mathrm{SINR}_k^{eff}(W, H_k, G_k, \sigma^2) = \Big({\prod\nolimits_{l \in \mathcal{L}_k} \mathrm{SINR}_l(W, H_k, g_l, \sigma^2) } \Big)^{\frac{1}{L_k}},\\
    &\mathrm{SINR}_l(W, H_k, g_l, \sigma^2) = \dfrac{| g_l H_k w_l  |^2}{\sum_{i \ne l}^{L} |   g_l H_k w_i  |^2 +  \sigma^2 \| g_l\|^2}, \quad \forall l \in \mathcal{L}_k,
\end{align}
where $w_l \in \mathbb C ^ T$ is a precoding for the $l$-th symbol, $g_l \in \mathbb C ^ {R_k}$ is a detection vector of the $l$-th symbol, $\sigma^2$ is a variance of Gaussian noise \eqref{eq:system_model}. The total power of transmitting antennas without loss of generality is assumed to be equal to one.

Additionally, we consider the function of \textit{Single-User SINR} (\textit{SUSINR}) :
\begin{equation}\label{SUSINR}
    \mathrm{SUSINR}(\widetilde{S}, \sigma^2) = \frac{1}{\sigma^2} \bigg(\prod_{k=1}^K \frac{1}{L_k} \bigg(\prod\nolimits_{l \in \mathcal{L}_k} s_l^2 \bigg)^{\frac{1}{L_k}}\bigg)^{\frac 1 K}.
\end{equation}
The formula (\ref{SUSINR}) describes channel quality for the specified user without taking into account the others. The matrix $\widetilde{ S} \in \mathbb C^{L \times L}$ contains all singular values of $\widetilde{S}_k$ on the main diagonal: $\widetilde{S} = \text{diag} (\widetilde{ S}_1 \dots \widetilde{S}_K) \in \mathbb C^{L \times L}$, and $\widetilde{S}_k \in \mathbb R ^ {L_k \times L_k}$ is a diagonal sub-matrix of $S_k$ consisting of $L_k$ largest singular values of the $k$-th user, and $s_l$ are the corresponding elements of $\widetilde{S}$ \cite{Conjugate}.

\subsection{Power Constraints}

It is important to note that each transmitting antenna has a restriction on power of the transmitter. Assuming all the symbols being transmitted are properly normalized, this results in the following constraints on the precoding matrix $W$:
\begin{equation}
    \|w^i\|^2 = \sum_k|w^i_k|^2\le\frac{1}{T}\ \ \forall i=1,\dots,T.
    \label{eq:power_constraints}
\end{equation}
The total power is assumed to be equal to one.
 
\subsection{Detection Matrices}

A detection matrix $G_k$ is specific to each user. Below we consider the main detection algorithms: MMSE and MMSE-IRC. In the case of MMSE, the detection matrix $G_k$ is calculated as \cite{MMSE}:
\begin{equation}
    G_k = (H_kW_k)^\mathrm{H}(H_kW_k(H_kW_k)^\mathrm{H}+\sigma^2 I)^{-1},
    \label{eq:MMSE}
\end{equation}
where matrix $H_kW_k$ is estimated using pilot signals on the user side.

\section{Spectral Efficiency Prediction Problem}\label{sec:SE Problem}

\subsection{Machine Learning Background}
In this subsection, we give a background on machine learning (ML) tasks and methods. In ML, we are given a training dataset $\mathcal{D}^{tr}=\{(x^{tr}_i, y^{tr}_i)\}_{i=1}^{N^{tr}}$ of $N^{tr}$ pairs (input object $x$, target $y$). The problem is to learn an algorithm $a(x)$ (also called prediction function) that predicts target $y$ for any new object $x$. Usually $x \in \mathbb{R}^d$, however, complex inputs could also be used. Conventionally used targets include $y \in \mathbb{R}$ (regression problem) or $y \in \{1, \dots, C\}$ (classification problem, $C$ is the number of classes). We are also given a set of $N^{te}$ testing objects, $\mathcal{D}^{te}=\{(x^{te}_i, y^{te}_i)\}_{i=1}^{N^{te}}$, and a metric $Q(a)=\sum_{i=1}^{N^{te}} q(y_i, a(x_i))$ that measures the quality of predicting target for new objects. This metric should satisfy the problem-specific requirements and could be non-differentiable. Summarizing, in order to use machine learning methods we need to specify objects, targets, metrics, and collect the dataset of (object, target) pairs.

\label{sec:se}
\subsection{Objects and Targets} Firstly, we consider \textbf{the problem of predicting the SE:} given $T$ antennas at the base station, $K$ users with $R_k$ antennas and $L_k$ symbols each, and set of channel matrices $\{H_k\}_{k=1}^K$ (input object), the problem is to predict the spectral efficiency SE \eqref{eq:spectral_efficiency} that could be achieved with some precoding algorithm (precoding is not modeled in this problem). We assume that $T=64$, $R_k=4$, $L_k=2 $ for all $k = 1 \dots K$ are fixed, while the number of users $K$ could be variable. Since $R_k$ is fixed, each object $\{H_k \in \mathbb{C}^{R_k \times T}\}_{k=1}^K$, which serves as an input for our ML algorithms, can be represented by three dimensional tensor $\mathcal{H} \in \mathbb{C}^{K \times 4 \times 64}$. 

\subsection{Metric} Since spectral efficiency prediction is a regression problem, we can use metrics known from this class of tasks. Namely, we use Mean Average Percentage Error (MAPE):
\begin{equation}\label{eq:MAPE}
    \mathrm{MAPE(a)} = \sum_{i=1}^{N^{te}} \biggl| \frac{\mathrm{SE}_i - a(\mathcal{H}_i)}{\mathrm{SE}_i} \biggr|,
\end{equation}
where $\mathrm{SE}_i$ is the spectral efficiency computed using specific precoding, and $a(\mathcal{H}_i)$ is  spectral efficiency predicted by the algorithm $a$ for the $i$-th object $\mathcal{H}_i$.

\subsection{Problem Statement and Research Questions}
Given the dataset $\{\mathcal{H}_i, \mathrm{SE}_i\}_{i=1}^{N^{tr}}$, our goal is to train an algorithm $a(\mathcal{H}_i)$ that predicts spectral efficiency SE based on the object $\mathcal{H}$. The algorithm may also take additional information as input, e. g. the level of noise $\sigma^2$ or SUSINR value. We also consider an alternative problem statement with user-wise spectral efficiency prediction. Our research questions are as follows: 
\begin{itemize}
    \item Can we predict the spectral efficiency (average or user-wise) with acceptable quality, e.\,g.\, with error on the test dataset, which is an order of magnitude less than the value we predict, i.\,e.\, $\mathrm{MAPE}<10\%$~\eqref{eq:MAPE}? 
    \item Which machine learning algorithm out of the considered classes (linear models, gradient boosting, neural networks) is the most suitable for working with channel data?
\end{itemize} 
    
\section{Proposed Methods}\label{sec:se_approach}

\subsection{Channel Dataset}

To obtain a dataset for this problem, we generate input channel matrices $H$, find precoding $W$ for each case using a certain precoding scheme, fixed for this particular dataset, and compute the target spectral efficiency SE using~\eqref{eq:spectral_efficiency}. To generate channel coefficients, we use Quadriga~\cite{Quadriga}, an open-source software for generating realistic radio channel matrices. We consider two Quadriga scenarios, namely Urban and Rural. For each scenario, we generate random sets of user positions and compute channel matrices for the obtained user configurations. We describe the generation process in detail in our work \cite{Conjugate}.

In the majority of the experiments, we consider three precoding algorithms: Maximum Ratio Transmission (MRT) \cite{MRT}, Zero Forcing (ZF) \cite{ZF} and LBFGS Optimization \cite{bobrov2021study}. The first two classic algorithms are quite simple, while the last one achieves higher spectral efficiency, however being slower. We  provide proof-of-concept results for Interference Rejection Combiner (IRC) SE \eqref{eq:IRC1} \cite{IRC}.

\subsection{Algorithm Features}

We consider well-known machine learning algorithms for regression task, namely linear models, gradient boosting, and fully connected neural networks. We rely on the following pipeline:
    \begin{enumerate}
        \item Take an object $\mathcal{H}$ as input (and possibly other inputs, e. g. SUSINR).
        \item Extract a feature vector representation $x = f(\mathcal{H})$ based on the raw input $\mathcal{H}$.
        \item Apply the algorithm $a$ to feature vector $x$ to obtain $a(x) = \mathrm{SE}$.
    \end{enumerate}
Further, we discuss how to obtain feature extraction procedure $f$ and prediction function $a$. Motivated by the exact formulas for the case of Maximum Ratio and Zero-Forcing precoding, we select the following \textit{default} features from the Singular Value Decomposition (SVD) of the $H = U^\mathrm{H} S V$ matrix, where $K$ is the number of users, and $L_i$ is the $i$-th user number of layers:
 \begin{itemize}
        \item $\{s^2_{ij}\}_{i,j=1}^{K, L_i}$ -- singular values;
        \item $\{c_{ijkm}\}_{\underset{i \ne j}{i,j,k,m=1}}^{K,K,L_i, L_j}$ - pairwise layer correlations, $c_{ijkm} = corr(V_{ik}, V_{jm}) = | V_{ik} V_{jm}|^2$.
    \end{itemize}
    
    
It can be shown that spectral efficiency~\eqref{eq:spectral_efficiency}, which is the value being predicted, depends only on the squares of the first $L_i$ singular values of $H_i$ of all users $i = 1 \dots K$ and the correlations between the first $L_i$ layers of all possible user pairs, which justifies this choice of \textit{default} features.

There are several issues with these features: 1) the number of features is variable, as the number of users $K$ varies among different objects $\mathcal{H} \in \mathbb{C}^{K \times 4 \times 64}$ in the dataset, while the aforementioned ML models take a feature vector of fixed size as input 2) the target objective, SE, is invariant to permutation of users, hence there should be a symmetry in feature representations. To address these issues, we propose two modifications of the \textit{default} features.

Firstly, we introduce symmetry with respect to user permutations by sorting the feature values along with user indices. For singular values, we sort users by the largest singular value. For pairwise correlations, we sort pairs of users by the largest correlation value of their layers.

Secondly, we alleviate the issue with a variable number of users by introducing \textit{polynomial} features. For a sequence $x = (x_1, x_2, \ldots, x_n)$ of features,  we can apply symmetric polynomial transformation, obtaining fixed number of features $poly_k(x) = (e_1(x), e_2(x), \ldots, e_k(x))$:
\[
    e_1(\{x_1, x_2, \ldots, x_n\}) = x_1 + \ldots + x_n
\]
\[
    e_2(\{x_1, x_2, \ldots, x_n\}) = \sum_{1 \leqslant j_1 < j_2 \leqslant n} x_{j_1} x_{j_2}  
\]
\[
   \dots
\]
\[
    e_k(\{x_1, x_2, \ldots, x_n\}) = \sum_{1 \leqslant j_1 < j_2 < \ \dotsm \ < j_k \leqslant n} x_{j_1} \dots x_{j_k}
\]

Therefore, in the following experiments, we consider 3 types of features: \textit{default}, \textit{sorted}, \textit{poly}$_{k}$. In some experiments, we also use a SUSINR as an additional feature. Models which are trained on \textit{default} and \textit{sorted} features require a dataset with a fixed number of users $K$=const, while \textit{poly}$_k$ features allow us to train models on a dataset with variable number of users $K$ between objects $\mathcal{H}_i$, which is a significant advantage. After extracting features, we use three ML algorithms: linear models, gradient boosting, and fully-connected neural networks. These algorithms are described in Section~\ref{sec:ml}.

\subsection{Machine Learning Methods Used for Predicting SE}\label{sec:ml}

In this work, we rely on the three most commonly used machine learning algorithms: linear models, gradient boosting, and neural networks. We discuss the positive and negative sides of these algorithms along with the description of how we apply them to 5G data. We compare all the results obtained by these models in Sec.~\ref{sec:experiments}.

\subsubsection{Linear Models}

We begin by evaluating simple \textit{linear regression models}. Linear models are very fast and memory-efficient and, being applied over a well-chosen feature space, could achieve high results in practice. Inspired by the fact that SUSINR is linear with respect to squares of singular values, we suppose that considering linear models is reasonable in our setting. We use the Scikit-learn implementation of linear regression and particularly, class \verb|LassoCV| which tunes the regularization hyperparameter using cross-validation.  

\textit{Linear models} are one of the simplest and fastest supervised learning algorithms. Linear models work with data represented in matrix form where each object $x$ is represented by a feature vector: $x \in \mathbb{R}^d$. The dimensionality $d$ is fixed and the same for all objects. A linear model predicts the target for the input object $x$ with a linear function $f(x, w) = \sum_{j=1}^d w_j x_j$. In this prediction function, \textit{parameters} $w$ (also called \textit{weights}) are unknown and need to be found based on the training data $\mathcal{D}^{tr}$. Usually, this is achieved by optimizing a loss function $\ell(y, a(x))$ that measures the penalty of predicting target $y$ with algorithm $f(x, w)$.

\subsubsection{Gradient Boosting}

We proceed by evaluating an out-of-the-box \textit{gradient boosting model}, a simple, but efficient nonlinear model. Gradient boosting is considered in as one of the best-performing algorithms for regression tasks. Moreover, gradient boosting is fast and does not require a lot of memory to store trained models. We use the CatBoost~\cite{Catboost} implementation of gradient boosting. The hyperparameters of our model are listed in Tab. \ref{tab:hyp_catboost_se_prediction}.

Gradient boosting processes input data and combines the predictions of several \textit{base} algorithms $b_1, \dots, b_N$, which are usually decision trees. The final prediction function is a weighted sum of base learners $a(x) = \sum_{n=1}^N \gamma_n b_n(x)$. Such algorithms are trained one by one, and each following algorithm tries to improve the prediction of the already built composition. 

Boosting models, as well as linear models, are suitable only for the data with a fixed feature vector dimensionality. That is always the case for \textit{poly}$_k$ features, while  \textit{default} and \textit{sorted} features require a fixed number of users $K$ in pairing.


\subsubsection{Neural Networks}
    
Finally, we consider \textit{fully-connected neural networks}. Their strength is the ability to capture complex nonlinear dependencies in data. In \textit{deep learning}, the prediction function $a$, called \textit{neural network}, is constructed as a composition of differentiable parametric functions: $a(x) = f(x, w)$ where \textit{parameters} $w \in \mathbb{R}^p$ are learned via the optimization of criteria $\ell(y, f(x, w))$.  The distinctive feature of neural networks is the wide range of architectures, i. e. different compositions $f(x, w)$, that can be chosen given the specifics of the particular problem.

However, the drawback of neural networks is that they require careful hyperparameter tuning, i. e. they usually achieve low results in out-of-the-box configuration. Thus, in practice, gradient boosting is often preferred over neural networks. Still, we conduct experiments using neural networks as well, to estimate their capabilities of processing channel data and obtaining reasonable prediction quality. One more drawback of neural networks is that they are slower than classic ML algorithms such as linear models or boosting, and require more memory to store parameters. But there are several techniques aimed at reducing the time and memory complexity of the trained models \cite{Sparcification, Sparcification2, Vardrop, Distillation}. 

Another important issue regarding neural networks is tuning their hyperparameters, which is essential for their performance. We train our fully-connected neural networks using SGD  with momentum as it performed better than another popular optimization algorithm, Adam \cite{kingma2014adam}. The mini-batch has a size 32. We choose the learning rate using a search on the grid $\{10^{-p}\}$, $p=2, \dots, 5$ minimizing the train mean squared error. This is how we ensure the network does not underfit, i. e. is capable of recognizing train objects. However, it could overfit, i. e. memorize train data without learning actual dependencies in the data. To avoid it, we utilize standard regularization techniques: weight decay and dropout. We tune their hyperparameters using grid search, minimizing MAPE on the held-out set.

For neural networks, we normalize input data, both features, and targets, as it is essential for training convergence. Specifically, for each feature and target, we compute the mean and standard deviation over the training data, subtract the mean values from the elements of all the feature vectors, and divide by the standard deviation. This procedure needs to be done for both training and testing data, with the mean and standard deviation being computed over training data. For computing metrics, we scale the target values back to the original scale.

\section{Experimental Results}\label{sec:experiments}


\begin{figure*}
\begin{center}
\includegraphics[height=8cm]{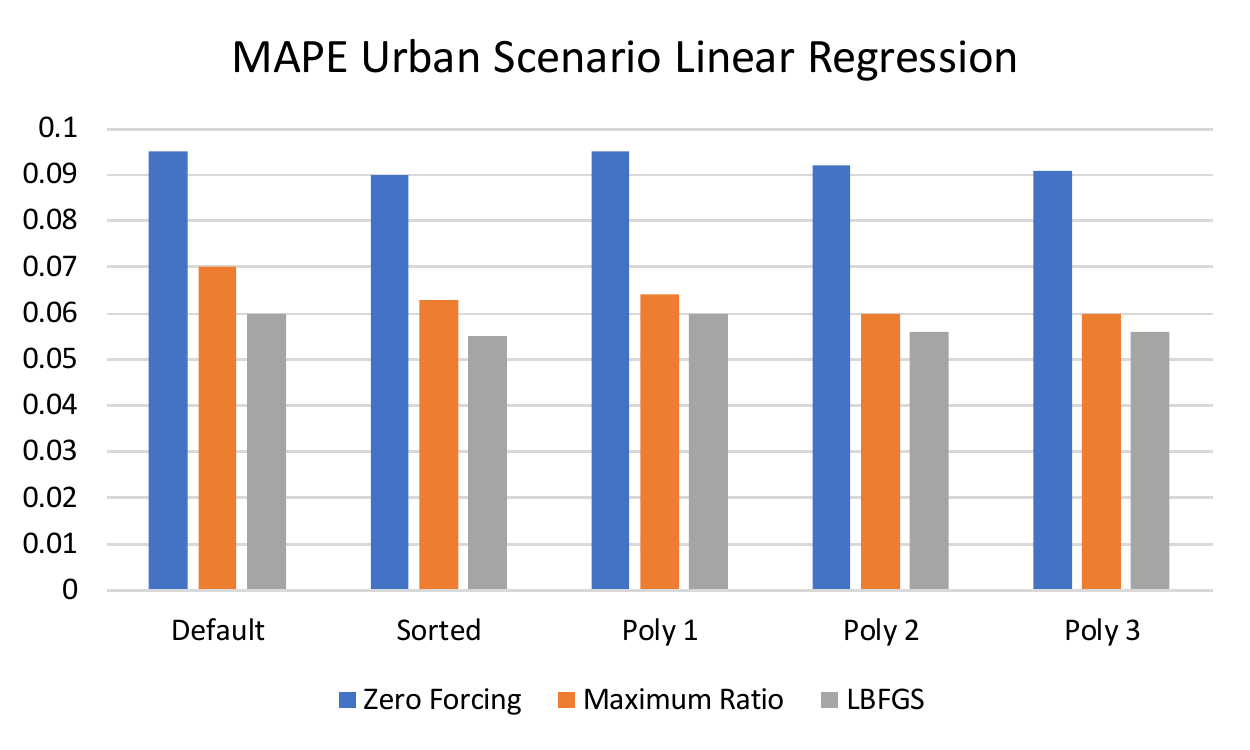}
\includegraphics[height=8cm]{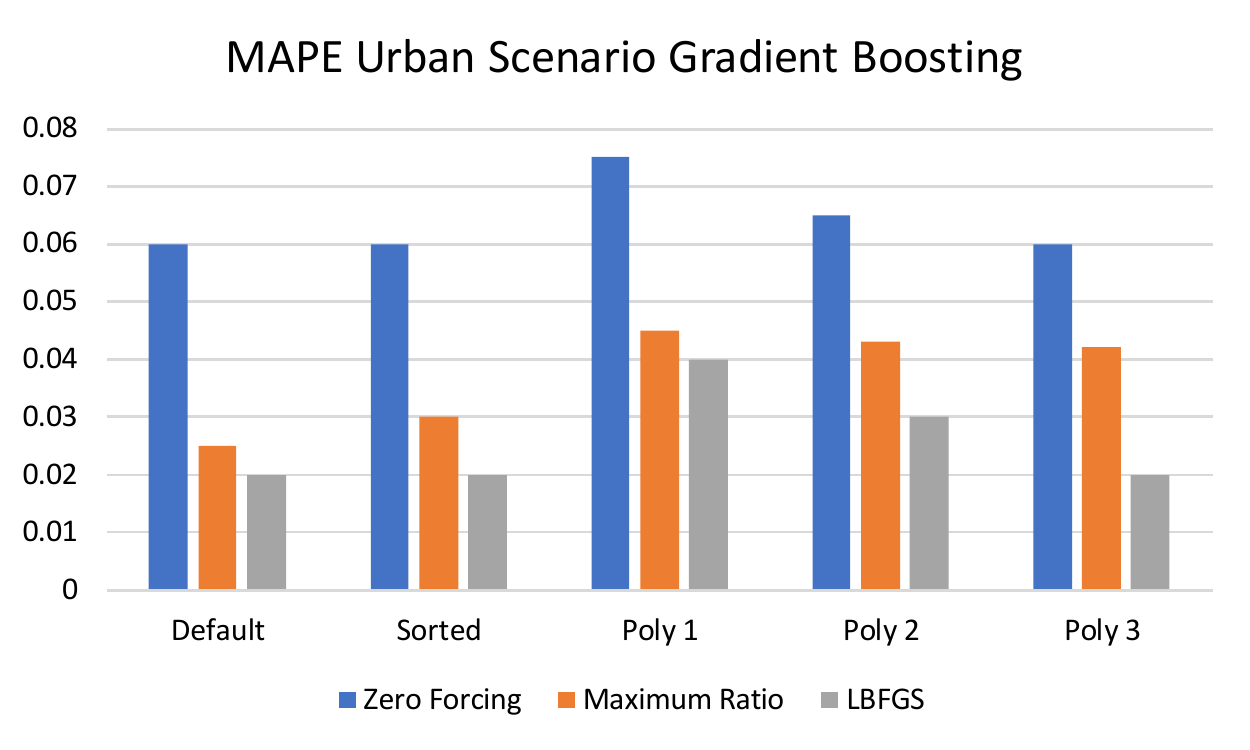}
\caption{Comparison  of  the  SE prediction algorithms: linear methods with different features and boosting with different features. The \textbf{urban} scenario is considered; and three different precoding methods -- Zero-Forcing (Blue), Maximum Ratio (Orange), LBFGS (Grey). The lower the MAPE value, the better.}
\label{fig:lin_boost_urban_setups}
\end{center}
\end{figure*}

\begin{figure*}
\begin{center}
\includegraphics[height=8cm]{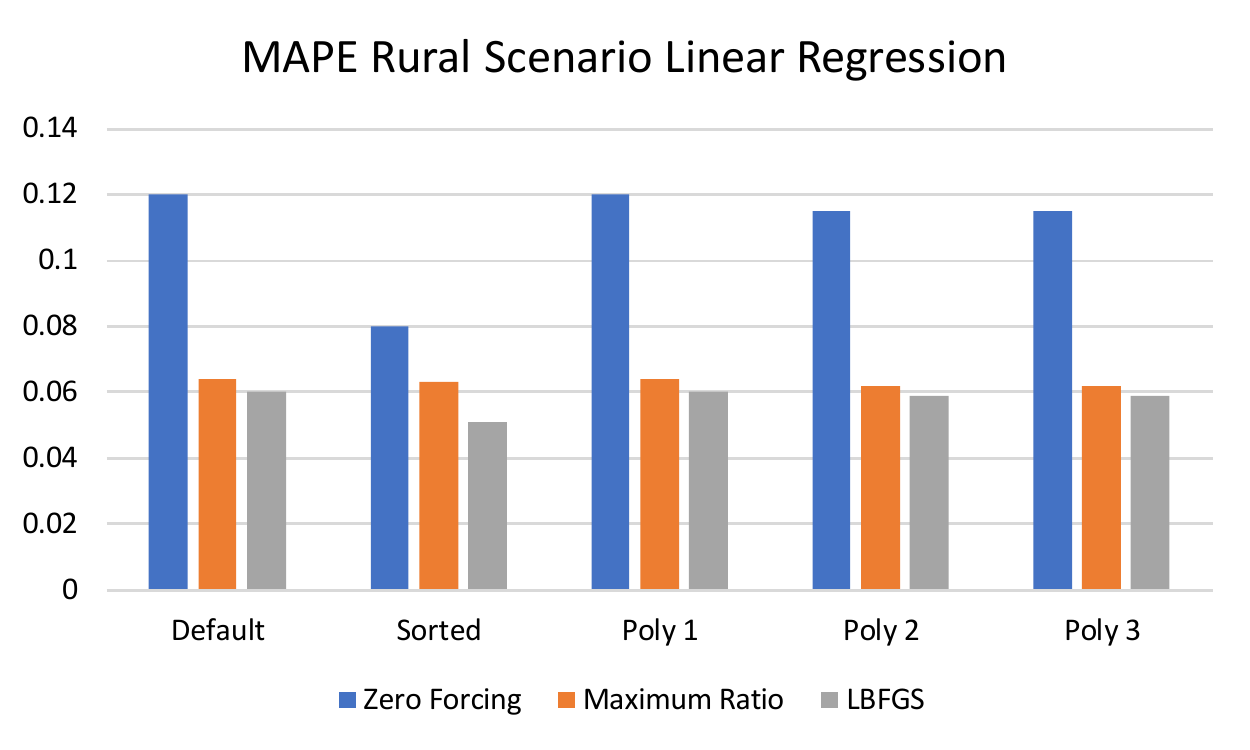}
\includegraphics[height=8cm]{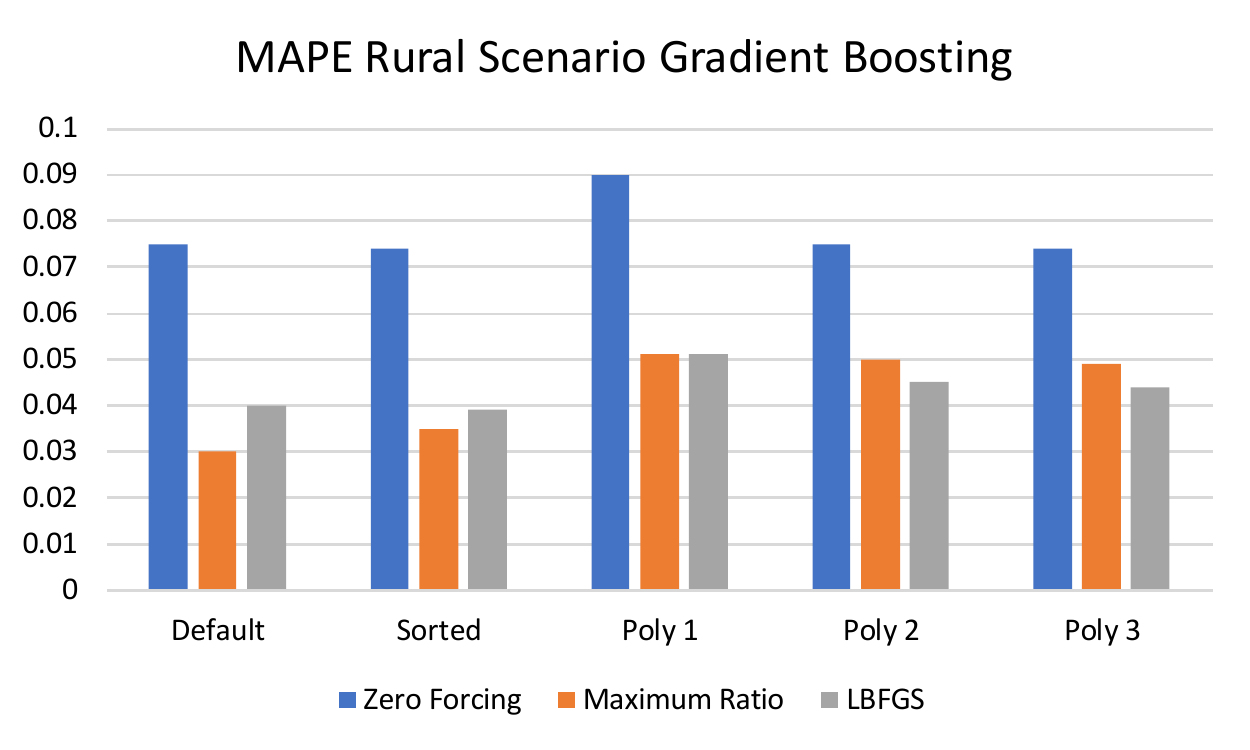}
\caption{Comparison  of  the  SE prediction algorithms: linear methods with different features and boosting with different features. The \textbf{rural} scenario is considered; and three different precoding methods -- Zero-Forcing (Blue), Maximum Ratio (Orange), LBFGS (Grey). The lower the MAPE value, the better.}
\label{fig:lin_boost_rural_setups}
\end{center}
\end{figure*}

\begin{figure*}
\begin{center}
\includegraphics[height=8cm]{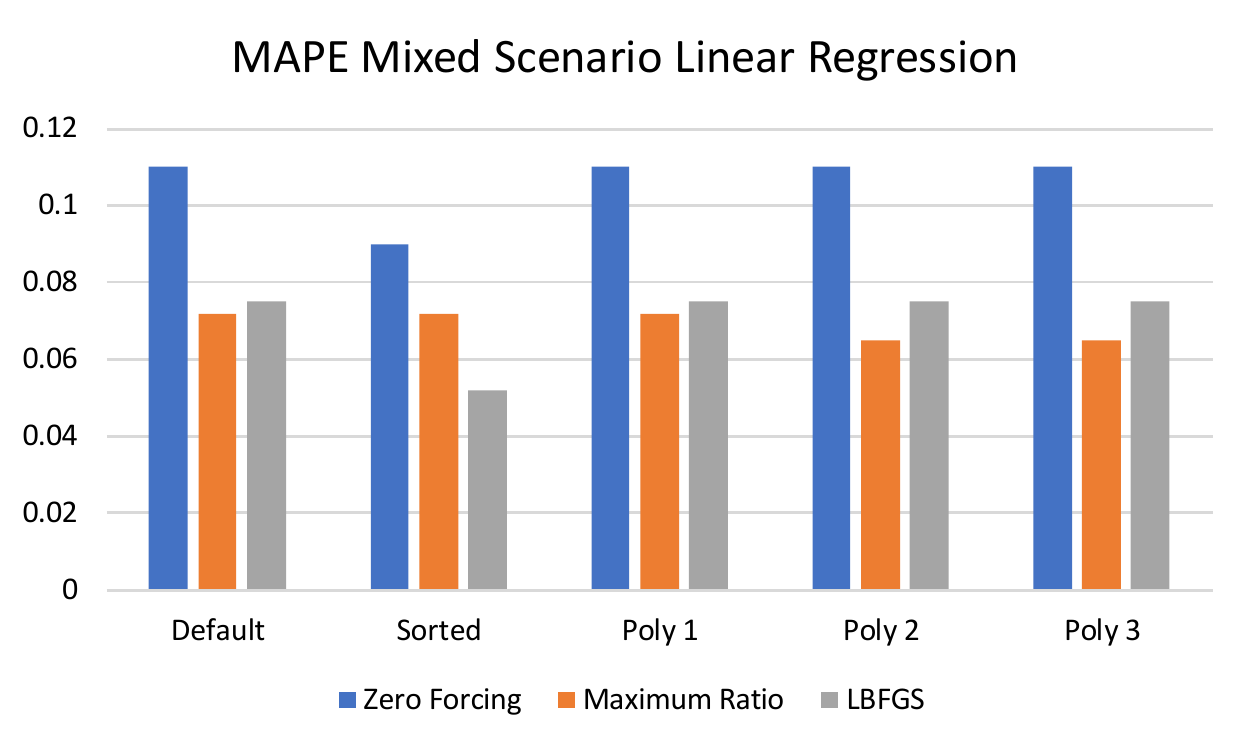}
\includegraphics[height=8cm]{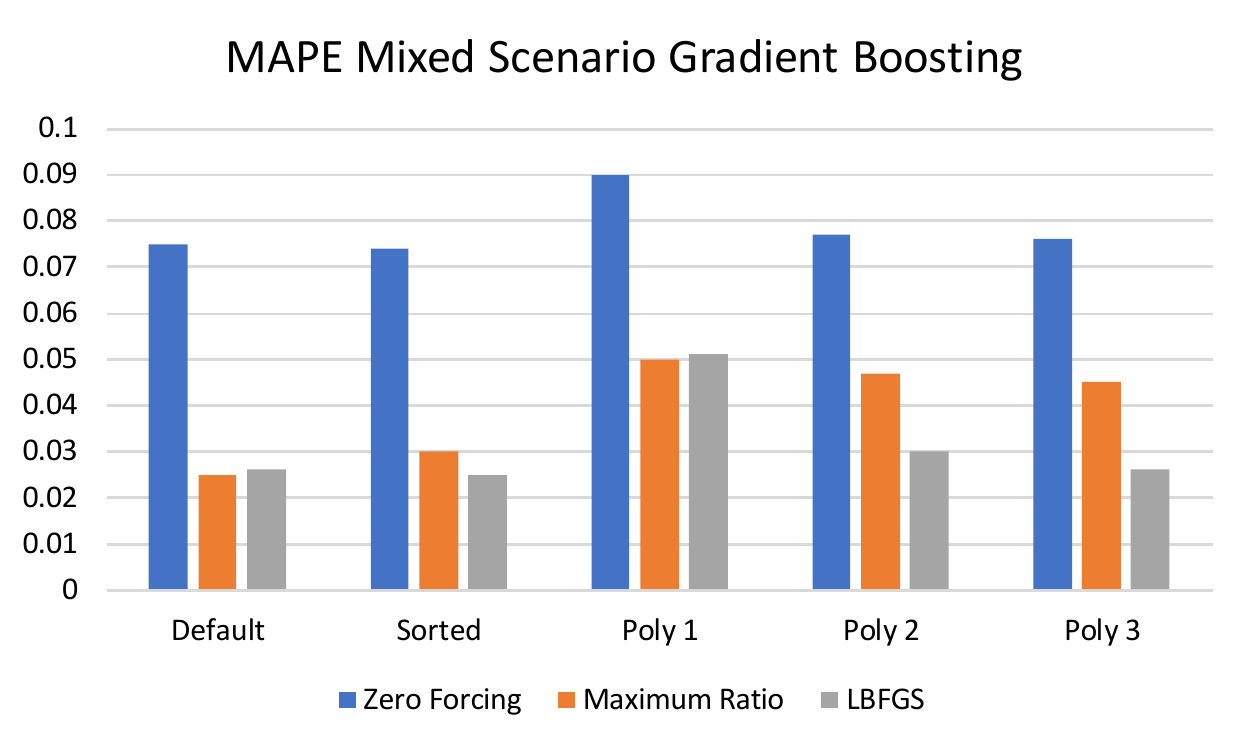}
\caption{Comparison  of  the  SE prediction algorithms: linear methods with different features and boosting with different features. The \textbf{mixed} scenario (urban and rural together) is considered; and three different precoding methods -- Zero-Forcing (Blue), Maximum Ratio (Orange), LBFGS (Grey). The lower the MAPE value, the better.}
\label{fig:lin_boost_mixed_setups}
\end{center}
\end{figure*}

This section contains the results for linear models, gradient boosting and neural networks. Firstly, we fix the number of users $K$ and analyze how our ML algorithms work for different precoding methods in different scenarios. Then we show how the results change for different values of $K$ and experiment with the solutions based on polynomial features in the case of variable number of users. We also measure time and memory complexity of the proposed algorithms. Finally, we apply the proposed approaches to the task of user-wise SE prediction.
In all of our experiments, train data size $N^{tr}=1.6 \cdot 10^4$ and test data size $N^{te}=3.6 \cdot 10^3$. 

The results for linear models and gradient boosting for different precoding methods and in different scenarios are shown in Figs.~(\ref{fig:lin_boost_urban_setups}, \ref{fig:lin_boost_rural_setups}, \ref{fig:lin_boost_mixed_setups}). Linear models demonstrate reasonable prediction quality. On the other hand, gradient boosting solutions provide significantly more accurate predictions. In terms of features, the best solutions are obtained using sorted features and polynomial features with the degree three. The results are only shown for a fixed number of users, $K=4$, but all the results hold for other considered values of $K$ (2 and 8 users).

\begin{figure*}
\begin{center}
        

\includegraphics[height=8cm]{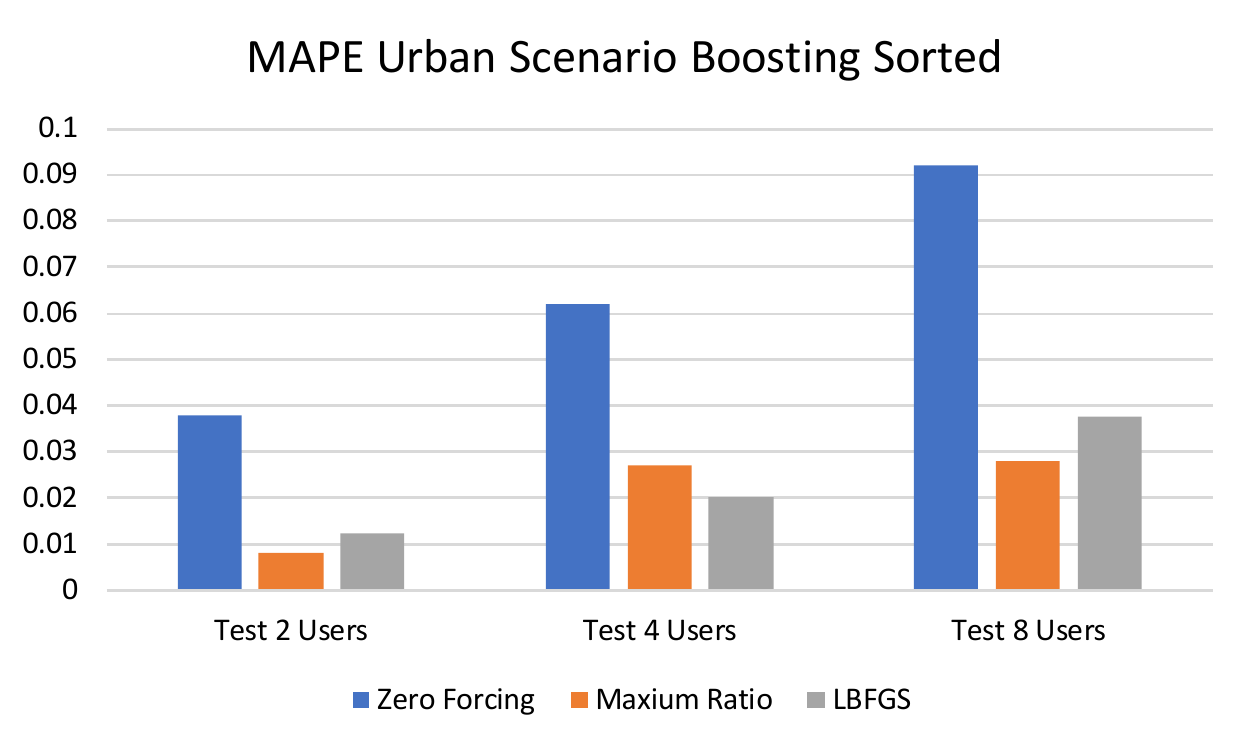}
\caption{Comparison  of  the  SE prediction algorithms  for different numbers of users -- 2, 4, and 8. All models are trained on sorted features datasets and a fixed number of users and validated on the test data with the same number of users. The lower the MAPE value, the better.}
\label{fig:diff_users}
\end{center}
\end{figure*}

\begin{figure*}
\begin{center}

\includegraphics[height=8cm]{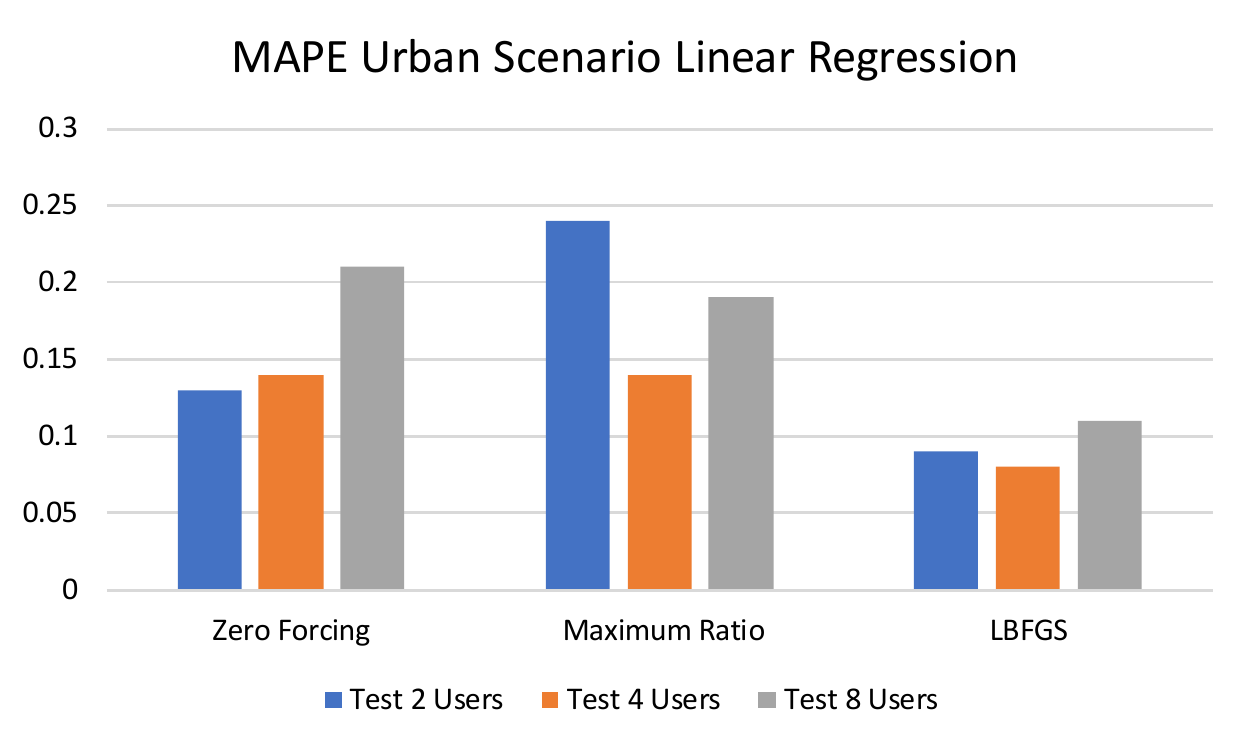}
\includegraphics[height=8cm]{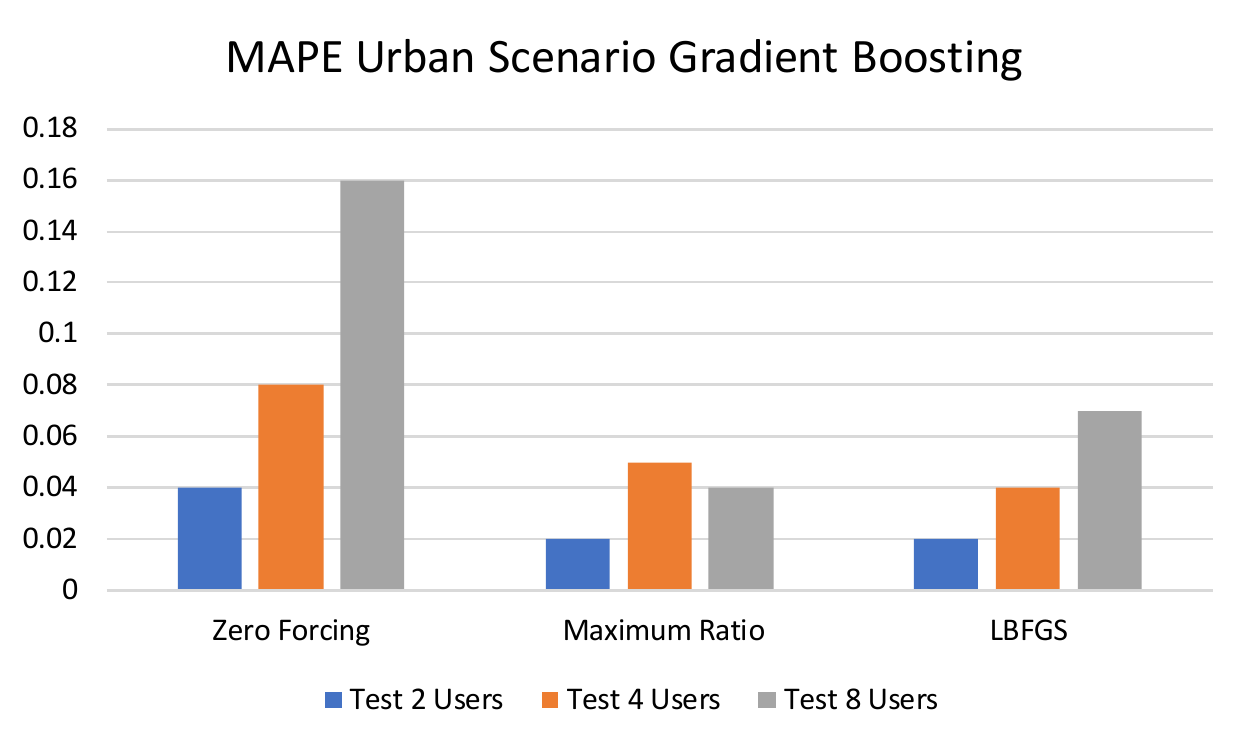} 

\caption{Comparison  of  the  SE prediction algorithms for different numbers of users -- 2, 4, and 8. All models are trained on a combined train dataset with the variable number of users (2, 4, and 8 together) and then tested on the test data with the fixed number of users. Polynomial features with the degree three are used. The lower the MAPE value, the better.}
\label{fig:all_users}
\end{center}
\end{figure*}

In Fig.~\ref{fig:diff_users}, we compare the results of the best-proposed models (boosting on sorted and polynomial features with the degree three) for different values of $K$. In this experiment, all models are trained on datasets with a fixed number of users and then are tested with the same number of users. The results show that the proposed algorithms perform well for all the considered numbers of users. 

The dimensionality of polynomial features is independent of the number of users $K$, therefore we can train all methods with polynomial features on the data with variable $K$. In Fig.~\ref{fig:all_users} we show the results of such an experiment. We train all models on a combined train dataset with a variable number of users (2, 4, and 8) and then test them on the test data with the fixed number of users. From the results we conclude that gradient boosting successfully tackles this problem for all precoding methods.

\subsection{Time and Memory Complexity}
\begin{table}
  \caption{Time complexity comparison. The average time for computing SE for one object is given in milliseconds. The computation time of the Zero-Forcing true SE values is given. Then, we compare the best-proposed models -- gradient boosting on sorted and polynomial features with the degree three. All these methods also use the same preprocessing, time for which is shown in the third column. The preprocessing consists of the computation of correlation matrix and squared singular values. }
  \label{time}
  \centering
  \vspace{0.2in}
  \begin{tabular}{c|c|ccc}
     \textbf{Number of} & \textbf{Zero-Forcing }& \textbf{Preprocessing}  &  \textbf{Boosting}  & \textbf{Boosting}  
 \\
\textbf{Users} & \textbf{ground truth}  & &  $+sorted$ \textbf{features} &  $+poly_3$ \textbf{features} \\
    \hline
     \{2\}  & 0.42 & 0.02 &  0.021 & 0.091
    \\
    \{4\}  & 0.78 & 0.05 &  0.023 & 0.092
    \\
    \{8\}  & 2.17 & 0.174 &  0.032 & 0.11
    \\
    \{2, 4, 8\} & 1.12 & 0.08 &  - & 0.097
    \\
  \end{tabular}
\end{table}

In Tab.~\ref{time}, we report the inference time of our best models, namely boosting algorithms with sorted and polynomial features, and of several reference algorithms, e. g. the computational time of true SE values for the Zero-Forcing precoding method. We observe that the time complexity of gradient boosting (including preprocessing) is an order smaller than the time of computing true SE values for the Zero-Forcing algorithm. 

\subsection{User-Wise SE Prediction}
In the previous experiments for SE prediction, we predicted SE for the whole pairing of users, i.e. averaged SE. Now, we wish to verify whether it is also possible to obtain reasonable prediction quality for each user separately since it can be useful in practice for selecting modulation coding scheme for each user \cite{bobrov2021massive}. We define the target SE$_u$ for each user as SE before averaging by the users: $\mathrm{SE}_k = \log_2(1 + \mathrm{SINR}^{eff}_k)$.

The features for user-wise SE prediction are the same as for the case of average SE prediction. For each user, we once again use the corresponding singular values, correlations between the chosen user's layers and the layers of other users, and additional features, such as noise power or equivalently SUSINR. These features characterize the relation of one particular user to all other users.  

\begin{figure*}
\begin{center}
\includegraphics[height=8cm]{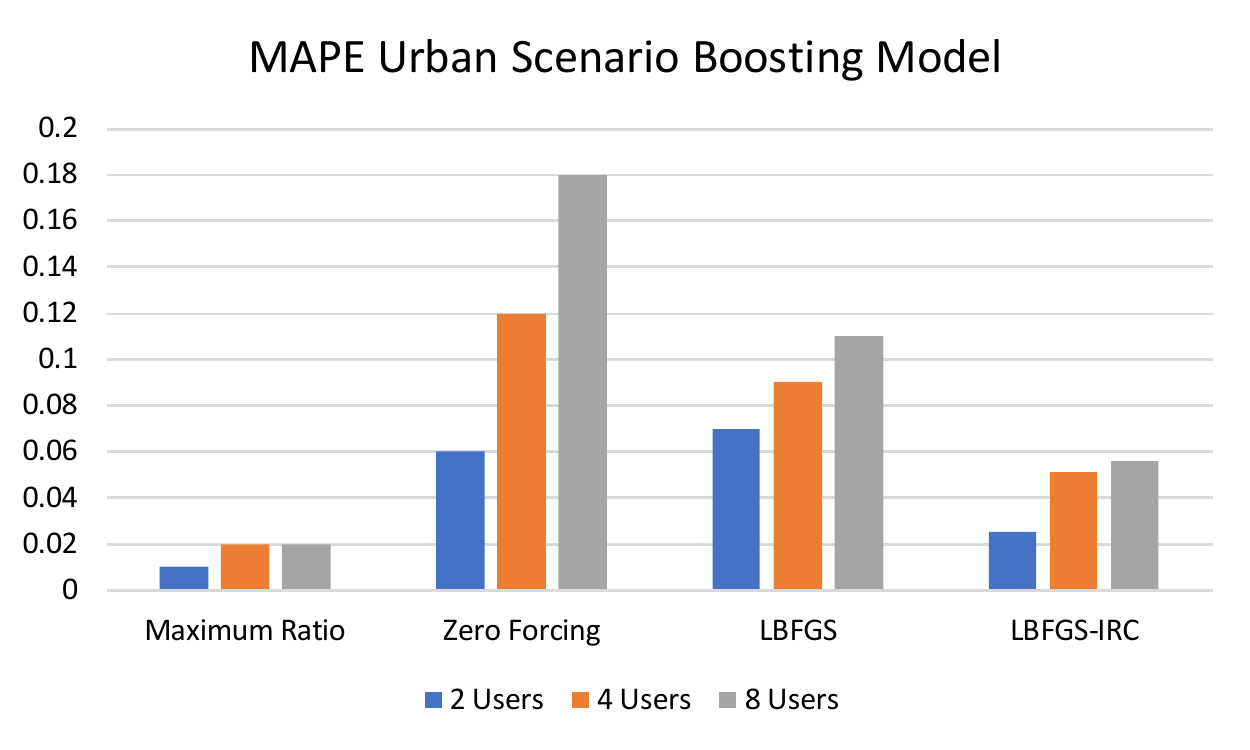}
\caption{Comparison  of  the  SE prediction algorithms  for user-wise SE prediction. The results for the Urban scenario, 2/4/8 users (separate models) are shown. Sorted features are used for boosting models. The lower the MAPE value, the better.}
\label{fig:per_user}
\end{center}
\end{figure*}

The results for the Urban scenario, 2/4/8 users (separate models) are shown in Fig.~\ref{fig:per_user}. Note that MAPE values in these experiments are not comparable to the values from previous sections, because of different target spaces. The results show that the gradient boosting approach outperforms the linear regression approach, for all considered user counts and precoding algorithms.

\subsection{SE-IRC Prediction} 
We also provide the results for \textit{SE-IRC}  \cite{IRC} prediction based on MMSE-IRC detection \eqref{eq:IRC1}, to verify that our method is robust to the change of the detection. For the case of MMSE-IRC, the detection matrix is calculated as:
\begin{equation}\label{eq:IRC1}
    G_k = \left( H_{k}W_{k} \right) ^\mathrm{H} \left( H_{k}W_{k} \left( H_{k}W_{k} \right) ^\mathrm{H} + R_{uu}^{k} + \sigma^2 I  \right) ^{-1},
\end{equation}
where the matrix  \(R_{uu}^{k} \) is related to unit symbol variance and is calculated as follows:
\begin{equation}\label{eq:IRC}
    R_{uu}^{k}=H_{k} \left(W W^\mathrm{H}-W_{k}W_{k}^\mathrm{H} \right) H_{k}^\mathrm{H}=H_{k} \left(  \sum _{u=1,u \neq k}^{K}W_{u}W_{u}^\mathrm{H} \right) H_{k}^\mathrm{H}.
\end{equation}
The results for the Urban scenario, 2/4/8 users (separate models) are shown in Fig.~\ref{fig:per_user}. From this results we conclude that the LBFGS-IRC composition of precoding and detection allows us to obtain better prediction quality with gradient boosting in comparison with other precoding schemes except MRT for all considered user counts.

\subsection{Results for Fully-Connected Neural Networks}
\begin{figure*}
\begin{center}

\includegraphics[height=8cm]{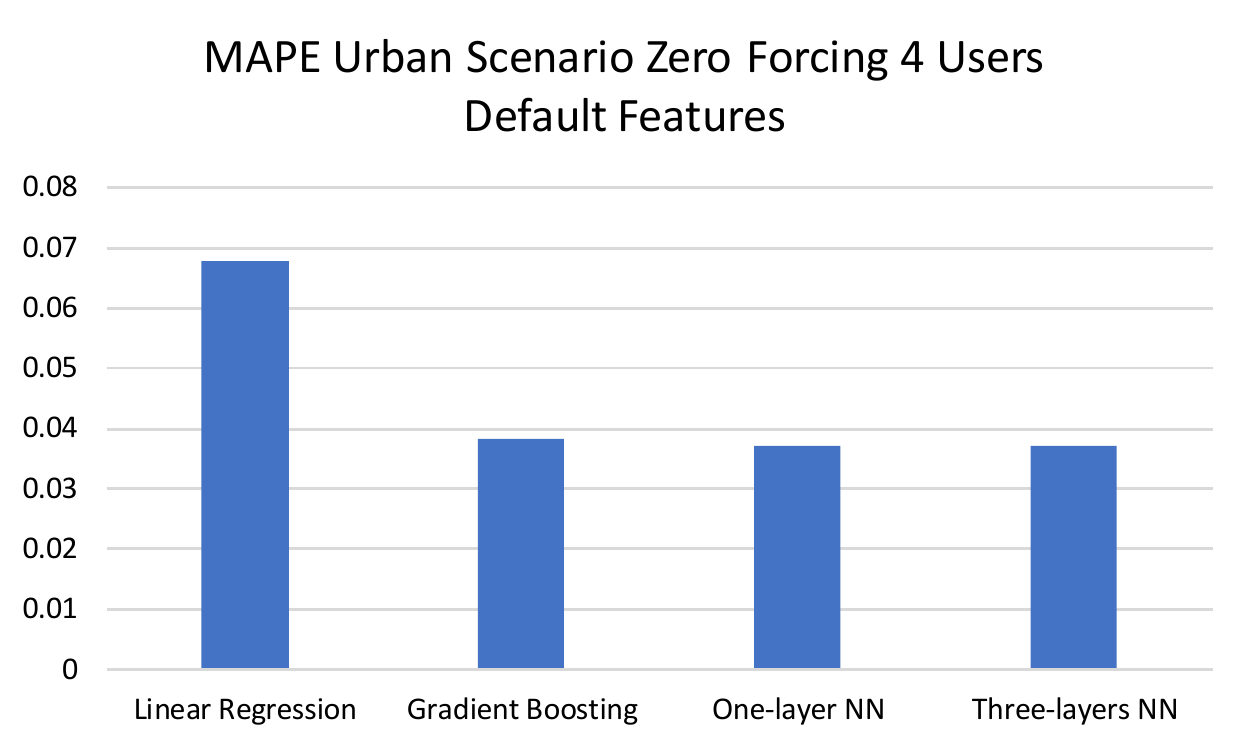}
\caption{Comparison  of  the  SE prediction algorithms: linear regression, gradient boosting, one-layer neural network (NN), three-layers NN with default features. One scenario is considered -- urban; and one precoding method -- Zero-Forcing. The lower the MAPE value, the better.}
\label{fig:fcn}
\end{center}
\end{figure*}

To test whether neural networks could improve feature-based spectral efficiency prediction, we train a fully-connected neural network on the \textit{default} features, for the Urban scenario, 4 users, Zero Forcing precoding method. We consider one and three hidden layers configurations with 200 neurons at each layer. Our results show that while linear models provide MAPE of 0.0679, and gradient boosting -- of 0.0383, the one-/three-layer neural networks achieve MAPE of  0.0371 / 0.0372 (see Fig. \ref{fig:fcn}). 

Thus, we conclude that using fully-connected neural networks is comparable to using gradient boosting in terms of prediction quality, while the tuning of network hyperparameters and the training procedure itself require significantly more time and memory than those of the other methods. However, these experiments showed that neural networks are in general applicable to channel data.

\section{Conclusion}\label{sec:conclusion}
To summarize, in this paper we consider the problem of spectral efficiency prediction using machine learning methods. We looked at three methods of forming feature vector representations for user channel data. We compared several machine learning algorithms, namely linear models, gradient boosting, and fully connected neural networks. We found that gradient boosting applied to sorted objects provides the best results, while linear models achieve lower quality. The neural networks perform similarly to boosting, but require more time and effort to set up. In almost all cases, prediction quality reaches MAPE below 10\% using gradient boosting and neural networks. This valuable result will allow us to significantly improve the quality of MIMO wireless communication in the future.

\section*{Acknowledgements}

Authors are grateful to Dmitry Kovkov and Irina Basieva for discussions.

\section*{Funding}

The work is funded by Huawei Technologies.

\bibliographystyle{tfs}
\bibliography{interacttfssample}

\begin{thebibliography}{10}
\providecommand{\MR}{\relax\unskip\space MR }
\providecommand{\url}[1]{\normalfont{#1}}
\providecommand{\urlprefix}{Available at }

\bibitem{alouini1999area}
M.S. Alouini and A.J. Goldsmith, \emph{Area spectral efficiency of cellular
  mobile radio systems}, IEEE Transactions on vehicular technology 48 (1999),
  pp. 1047--1066.

\bibitem{Benson06}
H.P. Benson, \emph{Maximizing the ratio of two convex functions over a convex
  set}, Naval Research Logistics (NRL) 53 (2006), pp. 309--317.

\bibitem{Conjugate}
E. Bobrov, B. Chinyaev, V. Kuznetsov, H. Lu, D. Minenkov, S. Troshin, D.
  Yudakov, and D. Zaev, \emph{{Adaptive Regularized Zero-Forcing Beamforming in
  Massive MIMO with Multi-Antenna Users}}  (2021).

\bibitem{bobrov2021massive}
E. Bobrov, D. Kropotov, and H. Lu, \emph{Massive {MIMO} adaptive modulation and
  coding using online deep learning algorithm}, IEEE Communications Letters
  (2021).

\bibitem{bobrov2021study}
E. Bobrov, D. Kropotov, S. Troshin, and D. Zaev, \emph{Finding better precoding
  in massive {MIMO} using optimization approach}, CoRR abs/2107.13440 (2021).
  \urlprefix\url{https://arxiv.org/abs/2107.13440}.

\bibitem{chataut2019channel}
R. Chataut and R. Akl, \emph{Channel Gain Based User Scheduling for 5G Massive
  MIMO Systems}, in \emph{2019 IEEE 16th International Conference on Smart
  Cities: Improving Quality of Life Using ICT \& IoT and AI (HONET-ICT)}. IEEE,
  2019, pp. 049--053.

\bibitem{d2019uplink}
C. D'Andrea, A. Zappone, S. Buzzi, and M. Debbah, \emph{Uplink power control in
  cell-free massive MIMO via deep learning}, in \emph{2019 IEEE 8th
  International Workshop on Computational Advances in Multi-Sensor Adaptive
  Processing (CAMSAP)}. IEEE, 2019, pp. 554--558.

\bibitem{farajidana20093gpp}
A. Farajidana, W. Chen, A. Damnjanovic, T. Yoo, D. Malladi, and C. Lott,
  \emph{3GPP LTE downlink system performance}, in \emph{GLOBECOM 2009-2009 IEEE
  Global Telecommunications Conference}. IEEE, 2009, pp. 1--6.

\bibitem{foschini1998limits}
G.J. Foschini and M.J. Gans, \emph{On limits of wireless communications in a
  fading environment when using multiple antennas}, Wireless personal
  communications 6 (1998), pp. 311--335.

\bibitem{Gotoh01}
J.Y. Gotoh and H. Konno, \emph{Maximization of the ratio of two convex
  quadratic functions over a polytope}, Computational Optimization and
  Applications 20 (2001), pp. 43--60.

\bibitem{guo2020regression}
Y. Guo, C. Hu, T. Peng, H. Wang, and X. Guo, \emph{Regression-based uplink
  interference identification and SINR prediction for 5G ultra-dense network},
  in \emph{ICC 2020-2020 IEEE International Conference on Communications
  (ICC)}. IEEE, 2020, pp. 1--6.

\bibitem{heliot2012energy}
F. H{\'e}liot, M.A. Imran, and R. Tafazolli, \emph{On the energy
  efficiency-spectral efficiency trade-off over the mimo rayleigh fading
  channel}, IEEE Transactions on Communications 60 (2012), pp. 1345--1356.

\bibitem{Distillation}
G. Hinton, O. Vinyals, and J. Dean, \emph{{Distilling the Knowledge in a Neural
  Network}} (2015).

\bibitem{huang20}
H. Huang, W. Xia, J. Xiong, J. Yang, G. Zheng, and X. Zhu, \emph{Unsupervised
  learning-based fast beamforming design for downlink {MIMO}}, IEEE Access 7
  (2018), pp. 7599--7605.

\bibitem{huh2012achieving}
H. Huh, G. Caire, H.C. Papadopoulos, and S.A. Ramprashad, \emph{Achieving
  {"Massive MIMO"} spectral efficiency with a not-so-large number of antennas},
  IEEE Transactions on Wireless Communications 11 (2012), pp. 3226--3239.

\bibitem{huh2011network}
H. Huh, A.M. Tulino, and G. Caire, \emph{Network {MIMO} with linear
  zero-forcing beamforming: Large system analysis, impact of channel
  estimation, and reduced-complexity scheduling}, IEEE Transactions on
  Information Theory 58 (2011), pp. 2911--2934.

\bibitem{Quadriga}
S. Jaeckel, L. Raschkowski, K. B{\"o}rner, and L. Thiele, \emph{{QuaDRiGa}: {A}
  {3-D} multi-cell channel model with time evolution for enabling virtual field
  trials}, IEEE Transactions on Antennas and Propagation 62 (2014), pp.
  3242--3256.

\bibitem{MMSE}
M. Joham, W. Utschick, and J.A. Nossek, \emph{Linear transmit processing in
  mimo communications systems}, IEEE Transactions on signal Processing 53
  (2005), pp. 2700--2712.

\bibitem{kingma2014adam}
D.P. Kingma and J. Ba, \emph{Adam: A method for stochastic optimization}, arXiv
  preprint arXiv:1412.6980  (2014).

\bibitem{liu2016efficient}
X. Liu and X. Wang, \emph{Efficient antenna selection and user scheduling in 5G
  massive MIMO-NOMA system}, in \emph{2016 IEEE 83rd Vehicular Technology
  Conference (VTC Spring)}. IEEE, 2016, pp. 1--5.

\bibitem{MRT}
T.K. Lo, \emph{Maximum ratio transmission} 2 (1999), pp. 1310--1314.

\bibitem{Sparcification}
E. Lobacheva, N. Chirkova, and D. Vetrov, \emph{Bayesian sparsification of
  recurrent neural networks}, arXiv preprint arXiv:1708.00077  (2017).

\bibitem{Sparcification2}
E. Lobacheva, N. Chirkova, and D. Vetrov, \emph{{Bayesian Sparsification of
  Gated Recurrent Neural Networks}} (2018).

\bibitem{Vardrop}
D. Molchanov, A. Ashukha, and D. Vetrov, \emph{{Variational Dropout Sparsifies
  Deep Neural Networks}} (2017).

\bibitem{RZF19}
L.D. Nguyen, H.D. Tuan, T.Q. Duong, and H.V. Poor, \emph{Multi-user regularized
  zero-forcing beamforming}, IEEE Transactions on Signal Processing 67 (2019),
  pp. 2839--2853.

\bibitem{RZF}
C.B. Peel, B.M. Hochwald, and A.L. Swindlehurst, \emph{A vector-perturbation
  technique for near-capacity multiantenna multiuser communication-part {I}:
  channel inversion and regularization}, IEEE Transactions on Communications 53
  (2005), pp. 195--202.

\bibitem{Catboost}
L. Prokhorenkova, G. Gusev, A. Vorobev, A.V. Dorogush, and A. Gulin,
  \emph{{CatBoost: Unbiased Boosting with Categorical Features}}, in
  \emph{Proceedings of the 32nd International Conference on Neural Information
  Processing Systems}, Red Hook, NY, USA. Curran Associates Inc., NIPS'18,
  2018, p. 6639–6649.

\bibitem{IRC}
B. Ren, Y. Wang, S. Sun, Y. Zhang, X. Dai, and K. Niu, \emph{Low-complexity
  mmse-irc algorithm for uplink massive mimo systems}, Electronics Letters 53
  (2017), pp. 972--974.

\bibitem{rozenblit2018machine}
O. Rozenblit, Y. Haddad, Y. Mirsky, and R. Azoulay, \emph{Machine learning
  methods for sir prediction in cellular networks}, Physical Communication 31
  (2018), pp. 239--253.

\bibitem{sanguinetti2018deep}
L. Sanguinetti, A. Zappone, and M. Debbah, \emph{Deep learning power allocation
  in massive MIMO}, in \emph{2018 52nd Asilomar conference on signals, systems,
  and computers}. IEEE, 2018, pp. 1257--1261.

\bibitem{BOpt}
J. Snoek, H. Larochelle, and R.P. Adams, \emph{Practical bayesian optimization
  of machine learning algorithms}, Advances in neural information processing
  systems 25 (2012).

\bibitem{sun2017learning}
H. Sun, X. Chen, Q. Shi, M. Hong, X. Fu, and N.D. Sidiropoulos, \emph{Learning
  to optimize: Training deep neural networks for wireless resource management},
  in \emph{2017 IEEE 18th International Workshop on Signal Processing Advances
  in Wireless Communications (SPAWC)}. IEEE, 2017, pp. 1--6.

\bibitem{telatar1999capacity}
E. Telatar, \emph{Capacity of multi-antenna {Gaussian} channels}, European
  transactions on telecommunications 10 (1999), pp. 585--595.

\bibitem{ullah2020machine}
R. Ullah, S.N.K. Marwat, A.M. Ahmad, S. Ahmed, A. Hafeez, T. Kamal, and M.
  Tufail, \emph{A machine learning approach for 5g sinr prediction},
  Electronics 9 (2020), p. 1660.

\bibitem{van2019sum}
T. Van~Chien, E. Bjornson, and E.G. Larsson, \emph{Sum spectral efficiency
  maximization in massive MIMO systems: Benefits from deep learning}, in
  \emph{ICC 2019-2019 IEEE International Conference on Communications (ICC)}.
  IEEE, 2019, pp. 1--6.

\bibitem{Transformer}
A. Vaswani, N. Shazeer, N. Parmar, J. Uszkoreit, L. Jones, A.N. Gomez, L.u.
  Kaiser, and I. Polosukhin, \emph{Attention is All you Need}, in
  \emph{{Advances in Neural Information Processing Systems}}, I. Guyon, U.V.
  Luxburg, S. Bengio, H. Wallach, R. Fergus, S. Vishwanathan, and R. Garnett,
  eds., Vol.~30. Curran Associates, Inc., 2017.

\bibitem{SE}
S. Verd{\'u}, \emph{Spectral efficiency in the wideband regime}, IEEE
  Transactions on Information Theory 48 (2002), pp. 1319--1343.

\bibitem{wang2013spectral}
D. Wang, J. Wang, X. You, Y. Wang, M. Chen, and X. Hou, \emph{Spectral
  efficiency of distributed {MIMO} systems}, IEEE Journal on Selected Areas in
  Communications 31 (2013), pp. 2112--2127.

\bibitem{RZF2}
Z. Wang and W. Chen, \emph{Regularized zero-forcing for multiantenna broadcast
  channels with user selection}, IEEE Wireless Communications Letters 1 (2012),
  pp. 129--132.

\bibitem{wannstrom2013lte}
J. Wannstrom, \emph{{LTE}-advanced}, Third Generation Partnership Project
  (3GPP)  (2013).

\bibitem{xia20}
W. Xia, G. Zheng, Y. Zhu, J. Zhang, J. Wang, and A.P. Petropulu, \emph{A deep
  learning framework for optimization of {MISO} downlink beamforming}, IEEE
  Transactions on Communications 68 (2019), pp. 1866--1880.

\bibitem{ZF}
T. Yoo and A. Goldsmith, \emph{On the optimality of multiantenna broadcast
  scheduling using zero-forcing beamforming}, IEEE Journal on selected areas in
  communications 24 (2006), pp. 528--541.

\bibitem{zappone2018model}
A. Zappone, M. Di~Renzo, M. Debbah, T.T. Lam, and X. Qian, \emph{Model-aided
  wireless artificial intelligence: Embedding expert knowledge in deep neural
  networks towards wireless systems optimization}, arXiv preprint
  arXiv:1808.01672  (2018).

\bibitem{zhao2020power}
Y. Zhao, I.G. Niemegeers, and S.H. De~Groot, \emph{Power allocation in
  cell-free massive mimo: A deep learning method}, IEEE Access 8 (2020), pp.
  87185--87200.

\end{thebibliography}

\section{Appendix. Transformer-Based Method}\label{sec:appendix}

\begin{figure*}
\begin{center}
\includegraphics[height=8cm]{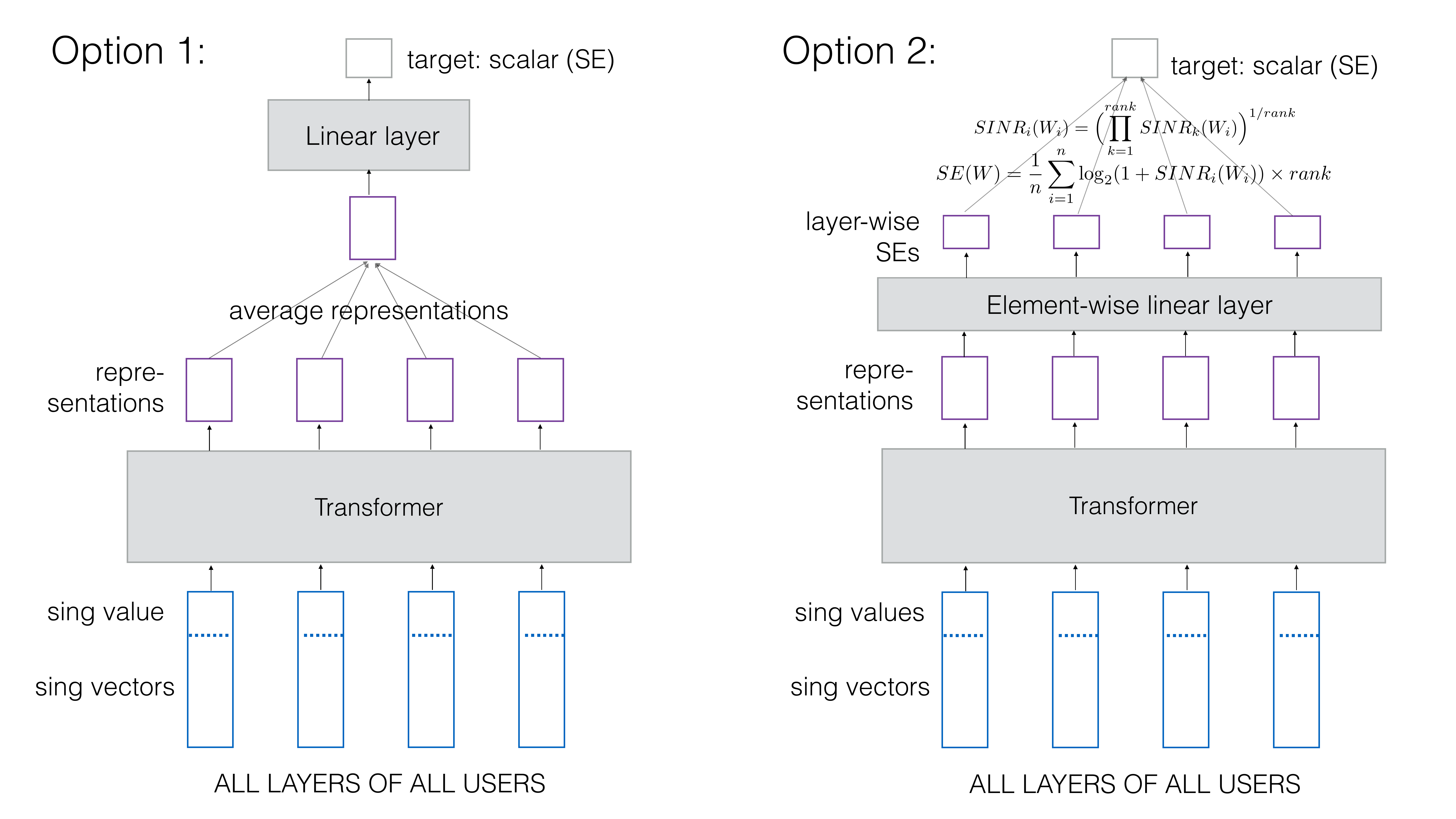}
\caption{Illustration of the transformer architecture for predicting spectral efficiency. Left: the direct prediction of the mean spectral efficiency. Right: the prediction of layer-wise spectral efficiencies.}
\label{fig:Tr_se_img}
\end{center}
\end{figure*}


In this section, we focus on a transformer state-of-the-art model, which has been successfully applied to the tasks of text, image, audio processing~\cite{Transformer}. The transformer model is well-suited for processing channel data since it supports the variable number of input elements (users/layers in our case). 

Our transformer-based architecture is illustrated in Fig.~\ref{fig:Tr_se_img}. It takes singular vectors and singular values of all layers of all users as input. Specifically, the transformer's input has a shape $(K \cdot L_k, 2 \cdot T + 1)$, where $K$ is the number of users in pairing, $L_k$ is $k$-th user number of symbols ($L_k=2, k=1\dots K$) and $T$ is the number of antennas at the base station ($T=64$). The shape $2 \cdot T + 1$ appears because the real and imaginary parts of $T$-sized singular vectors, and also a singular value are combined in one layer representation. After applying the transformer, the output has a shape $(K \cdot L_k, d_{\mathrm{model}})$, where $d_{\mathrm{model}}$ is the hyperparameter. We then apply fully-connected layer to each of the $K \cdot L_k$ $d_\mathrm{model}$-dimensional transformer outputs and obtain the output of a shape $(K \cdot L_k, 1)$. 

We consider two approaches in our experiments. The first option is to apply an arbitrary simple transformation, e. g. averaging over $K \cdot L_k$ values, to map the matrix to the scalar and then train the model so that this scalar approximates spectral efficiency. The second option is to use additional supervision. 

When we generate data, there is an intermediate step where we compute layer-wise spectral efficiencies: $\mathrm{SE}_l = \log_2(1 + \mathrm{SINR}_l)$, so that they are later averaged to obtain the mean spectral efficiency~\eqref{eq:spectral_efficiency}. Thus, we can train our transformer so that it would predict these layer-wise spectral efficiencies of shape $(K \cdot L_k, 1)$. To obtain the final spectral efficiency~\eqref{eq:spectral_efficiency} we use simple averaging.

When transformers are applied to sequences, positional encodings are often used to encode the order of the elements. Since our input data is order-free, we do not use positional encodings. We train our transformer model using the Adam optimization algorithm. We use the standard transformer's dropout with dropout rate selected from $\{0, 0.1, 0.2\}$ based on the performance on the held-out dataset. Hyperparameters of the model were selected using Bayesian optimization \cite{BOpt}.

Same as for fully-connected neural networks, we also use data normalization for transformers. Using proper normalization is essential for neural networks, otherwise, training would be unstable or even diverge. For normalization purposes mean and standard deviation values are computed over all layers of all users. The procedure is the same as for fully-connected networks. 

\begin{figure*}
\begin{center}
\includegraphics[height=8cm]{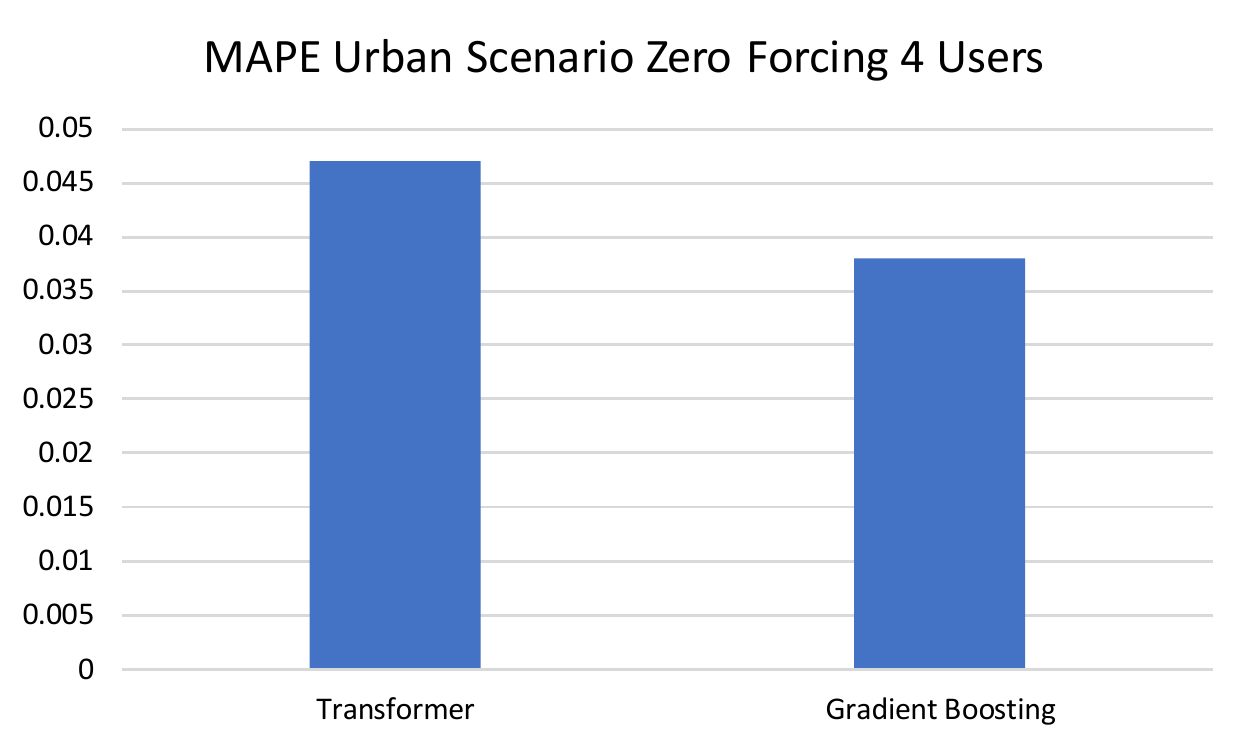}
\caption{Comparison  of  the  SE prediction algorithms: transformer and gradient boosting. Gradient boosting is trained on sorted features. The considered scenario is Urban; and the precoding method used is Zero-Forcing. The number of users is 4. The lower the MAPE value, the better.}
\label{fig:transformer_boosting}
\end{center}
\end{figure*}

In the experiments in Fig. \ref{fig:transformer_boosting} the transformer has 3 layers, $d_{model}=64$, the scenario is Urban, the number of users is 4, Zero Forcing precoding method is used. With proper data normalization, the transformer achieved MAPE of 0.047, and gradient boosting achieved 0.038. We also experimented with larger transformers, and they did not achieve lower error.

\begin{table}
        \centering
        \begin{tabular}{ll}
            \toprule
            \midrule
            auto\_class\_weights                &            None \\
            bayesian\_matrix\_reg               &             0.1 \\
            best\_model\_min\_trees              &               1 \\
            boost\_from\_average                &            True \\
            boosting\_type                     &           Plain \\
            bootstrap\_type                    &             MVS \\
            border\_count                      &             254 \\
            classes\_count                     &               0 \\
            depth                             &               6 \\
            eval\_metric                       &            MAPE \\
            feature\_border\_type               &    GreedyLogSum \\
            grow\_policy                       &   SymmetricTree \\
            iterations                        &            1000 \\
            l2\_leaf\_reg                       &               3 \\
            leaf\_estimation\_backtracking      &  AnyImprovement \\
            leaf\_estimation\_iterations        &               1 \\
            leaf\_estimation\_method            &           Exact \\
            learning\_rate                     &            0.03 \\
            loss\_function                     &             MAE \\
            max\_leaves                        &              64 \\
            min\_data\_in\_leaf                  &               1 \\
            model\_shrink\_mode                 &        Constant \\
            model\_shrink\_rate                 &               0 \\
            model\_size\_reg                    &             0.5 \\
            nan\_mode                          &             Min \\
            penalties\_coefficient             &               1 \\
            posterior\_sampling                &           False \\
            random\_seed                       &             228 \\
            random\_strength                   &               1 \\
            rsm                               &               1 \\
            sampling\_frequency                &         PerTree \\
            score\_function                    &          Cosine \\
            sparse\_features\_conflict\_fraction &               0 \\
            subsample                         &             0.8 \\
            task\_type                         &             CPU \\
            use\_best\_model                    &           False \\
            \bottomrule
            \end{tabular}
        \caption{Hyperparameters of the default CatBoost model for SE prediction.}
        \label{tab:hyp_catboost_se_prediction}
    \end{table}

\end{document}